\documentclass[amsmath,amssymb,10pt,aps,prb,twocolumn,floatfix,superscriptaddress]{revtex4-2}
\usepackage{graphicx}
\usepackage[dvipsnames]{xcolor}
\usepackage[colorlinks=true,linktoc=page,linkcolor=blue,citecolor=blue,urlcolor=blue]{hyperref}
\usepackage[capitalise]{cleveref}
\usepackage{bm}
\usepackage{breqn}
\usepackage{physics}
\usepackage{mathrsfs} 
\usepackage{mathtools}
\usepackage{chngcntr}

\def\K{\ensuremath{\text{K}}}
\def\Kp{\ensuremath{\text{K}'}}
\makeatletter
\let\cat@comma@active\@empty
\makeatother
\begin{document}
\title{Josephson junction of minimally twisted bilayer graphene}

\author{Ritajit Kundu}
\email{ritajit@iitk.ac.in}
\affiliation{Department of Physics, Indian Institute of Technology Kanpur, Kanpur 208016, India}
\author{Arijit Kundu}
\email{kundua@iitk.ac.in}
\affiliation{Department of Physics, Indian Institute of Technology Kanpur, Kanpur 208016, India}

\begin{abstract}
We theoretically investigate the transport properties of a Josephson junction composed of superconductor/minimally twisted bilayer graphene/superconductor structures. In the presence of an out-of-plane electric field, the low-energy physics is best described by a network of domain-wall states. Depending on system parameters, they lead to the emergence of zig-zag or pseudo-Landau level modes with distinct transport characteristics. Specifically, we find zig-zag modes feature linear dispersion of Andreev bound states, resulting in a $4\pi$-periodic Josephson current. In contrast, pseudo-Landau level modes exhibit flat Andreev bound states and, consequently, a vanishing bulk Josephson current. Interestingly, edge states can give rise to $4\pi$-periodic Josephson response in the pseudo-Landau level regime. We also discuss experimental signatures of such responses.
\end{abstract} 

\maketitle 
\section{Introduction} 

The discovery of correlated insulating states and unconventional superconductivity in twisted bilayer graphene (TBG) \cite{Cao_2018_1,Cao_2018} has generated significant interest in moiré materials. Such systems are often described by a periodic potential induced by the interference pattern of two rotationally misaligned graphene layers \cite{Bistritzer_2011}. Moiré materials exist at the intersection of two paradigms: topological and electronic correlation physics \cite{andrei2020graphene,
andrei2021marvels,
adak2024tunable,
nuckolls2024microscopic}. The coexistence of these effects promotes novel electronic phases that did not exist in each of the paradigms individually. Such as, around the magic angle of 1$^\circ$, twisted bilayer graphene hosts flat bands, where electron correlation effects dominate, giving rise to various broken symmetry phases \cite{yankowitz2019tuning,Xie_2021,Sharpe_2019,Cao_2020,
lu2019superconductors,Polshyn_2019,liu2021tuning,xie2019spectroscopic,choi2019electronic,nuckolls2020strongly,saito2021hofstadter,das2021symmetry,wu2021chern,potasz2021exact,kang2020non,hejazi2021hybrid}, whereas away from the magic angle, band topological properties such as the Chern number and geometric quantities such as the quantum metric may play an important role \cite{Abouelkomsan_2023,Hu_2019,Chen_2024,Wu_2020}.

In a minimally twisted (twist angle $\theta \ll 1^{\circ}$) bilayer graphene (MTBG), lattice relaxations become significant, leading to the formation of sharply defined triangular domains of alternating AB and BA Bernal stacked graphene \cite{Huang_2018,Yoo_2019} (refer to Fig. 1 of \cite{Yoo_2019} for a visualization of the system under consideration). The size of these triangular domains is determined by the moiré length scale, $l \sim a_0 / \theta[\text{rad}]$ ($a_0 = 2.46\,\text{\AA}$ is graphene's lattice constant). As $l \gg a_0$, the number of carbon atoms within these domains is of the order of $10^4 (\theta^\circ)^{-2}$, making atomistic quantum transport calculations challenging. 
However, in the presence of an electrostatic potential bias between the layers, these domains are insulating at charge neutrality, and non-trivial valley Chern indices result in domain-wall (DW) modes per spin and valley \cite{Yin_2016,Killi_2010,zhang_valley_2013}. At small temperatures, the electronic transport properties of this system are then described by a network of such DW modes. A Chalker-Coddington network model \cite{chalker1988percolation} captures the low-energy electronic physics effectively, which is demonstrated in recent transport studies in MTBG \cite{de_beule_aharonov-bohm_2020,vakhtel_bloch_2022}. This model successfully explains Aharanov-Bohm oscillations and incorporates \emph{zig-zag} (ZZ) modes predicted from microscopic calculations \cite{Fleischmann_2019,Tsim_2020}. Moreover, under specific network parameters, it predicts circulating modes, termed \emph{pseudo-Landau level} (PLL) modes. The network becomes transparent in the presence of zig-zag modes, whereas the pseudo-Landau level modes render it insulating. Aharanov-Bohm oscillation in the presence of a magnetic field due to such a network of domain-wall modes has also been observed recently \cite{Rickhaus_2018,Xu_2019}.

Apart from magneto-conductance, it is also important to study the transport phenomena in MTBG in other experimental settings and how they arise from and differ among zig-zag or pseudo-Landau level modes. Such a study will provide us with additional insight into the microscopic underpinnings of these systems. In this regard, we choose a Josephson junction composed of superconductor/minimally twisted bilayer graphene/superconductor structures. Josephson junctions have been extensively studied near the magic angle in TBG \cite{de_Vries_2021,Rodan_Legrain_2021,D_ez_M_rida_2023,Alvarado_2023}. Carrier concentration can be tuned by electrostatic gates in these systems. With this gate control, it is possible to vary the local filling factors and have two superconducting regions separated by a non-superconducting regions within a single sample of TBG. 

For the MTBG, we adopt a phenomenological Chalker-Coddington network model from previous transport studies \cite{de_beule_aharonov-bohm_2020,vakhtel_bloch_2022} and the superconducting leads are considered to be $s$-wave superconductors. Although the microscopic origin of the network parameters remains unknown, the merit of the network model lies in the fact that it conforms to the microscopic symmetry of MTBG.
We perform a comprehensive study of Andreev bound states (ABS) and Josephson current within this junction for a range of network parameters. We show that zig-zag modes yield zero-energy Andreev bound states and $4\pi$ periodic Josephson current. Conversely, the pseudo-Landau level modes host perfectly flat Andreev bound states and a vanishing Josephson current. By tuning the network parameter as we tune from zig-zag modes to the pseudo-Landau level modes, the ABS gap out by merging two Dirac cones in the momentum space. For other network parameters, the Josephson current phase relation is either $2\pi$ or $4\pi$ periodic, or a combination of both.
Furthermore, we study the effect of edges in the Josephson junction and find that when pseudo-Landau level modes are present, the Josephson current becomes finite and is mediated through the network's edges only.

The remaining part of the paper is structured as follows: In Section \ref{sec:model}, we describe the network model and the Josephson junction; in Section \ref{sec:abs}, we outline the calculation procedure for Andreev bound states and Josephson current within a bulk Josephson junction; and in Section \ref{sec:numerics}, we present numerical results for the same. In Section \ref{sec:edge}, we extend our analysis to investigate the effects of edges in MTBG on Andreev bound states and Josephson current. Finally, we conclude with Section \ref{sec:discussion}.
\section{Josephson junction} 
\label{sec:model} 
\begin{figure*}[ht] \centering
	\includegraphics[width=\linewidth]{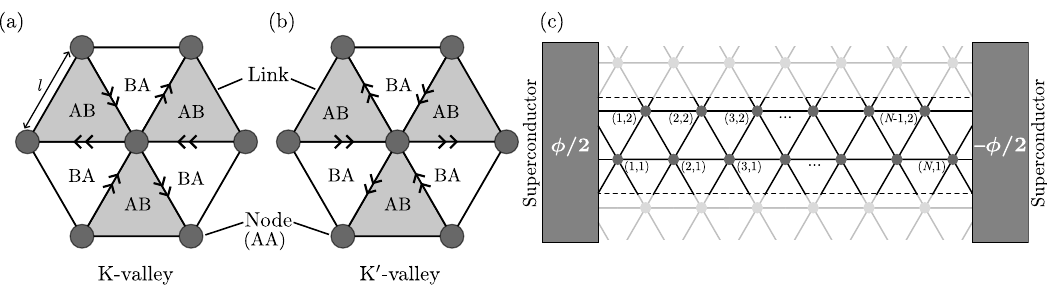} \caption{ 
Schematic representation of minimally twisted bilayer graphene (MTBG) for (a) K and (b) K\('\)-valley. The gray and white triangles denote AB and BA stacked regions of graphene, respectively. Their boundaries host two domain-wall modes per valley and spin. The velocities of these states are opposite for K and K\('\)-valleys. The dark circles represent the scattering nodes consisting of AA stacked regions. (c) shows the geometry considered here for the Josephson junction, composed of two $s$-wave superconducting leads (gray rectangles) connected by MTBG. Nodes are indexed by two integers $(m,n)$, which are connected by network links (black lines). We assume a periodic boundary condition in the transverse direction. The dashed lines encapsulate the unit-cell of the network.
  } \label{fig:network}
	\end{figure*}
We consider a Josephson junction (JJ) composed of two $s$-wave superconducting leads sandwiched by the MTBG. Superconducting order parameters for the left and
right leads are \( \Delta e^{\pm i \phi / 2} \), respectively. In this context, \(
\Delta \) represents the superconducting gap, while \(\phi\) refers to the
phase. We describe the MTBG utilizing a phenomenological network model~\cite{de_beule_aharonov-bohm_2020,De_Beule_2021_1,De_Beule_2021_2,vakhtel_bloch_2022,Wittig_2023}, which is an effective description of two minimally
twisted (\( \theta \ll 1^\circ \)) layers of graphene in the presence of an
out-of-plane electric field. For such small twist-angles, lattice relaxation leads to the
formation of triangular domains of AB and BA stacked graphene of moiré
length scale \( l = a_0 / \theta[\text{rad}] \). Fig.~\ref{fig:network}
(a) and (b) depict the schematic diagram of the triangular domains of AB
and BA bilayer graphene for the K and K$'$ valleys, respectively. 
In the presence of an electrostatic potential (\(U\)), the energy spectrum of the bulk of the
domains becomes gapped, leaving ballistic domain-wall (DW) modes on the domain
boundaries. There are two DW modes per valley per spin, as the change of
valley-Chern number across the AB and BA domain is \( \Delta C_{\text{v}} =
\pm 2\, \mathrm{sgn}(U / t_{\perp}) \) \cite{Killi_2010,zhang_valley_2013,Ju_2015}, where \(\pm\) stands for AB and BA domains respectively and \( t_{\perp} \) is the coupling constant between A and B sublattice of the two layers. The propagation
directions of modes in different valleys are reversed as the time-reversal symmetry is intact in MTBG. These DW modes form the \emph{links} of the network. Electrons propagate freely on the domain boundaries for a duration of \( \tau = l / v_F \) (\( v_F \) is the Fermi velocity of the DW modes) before they all come to AA stacked regions and scatter among
themselves (see Fig.~\ref{fig:network} (a), (b)). The AA stacked regions, which remain gapless even when an electric field is applied, are the \emph{nodes} of the
scattering network (see Fig.~\ref{fig:network} (c)). As the
scattering regions are smooth with respect to the atomic scale, scattering between the graphene valleys can be neglected.
Every node has three incoming and three outgoing channels, where each channel
corresponds to \emph{two} DW modes per spin and valley. Therefore, the scattering
matrix \( S \) of each node is a \( 6 \times 6 \) unitary matrix. The
scattering matrix acts on incoming modes and returns outgoing modes: \(
\bm b_{\text{out}}\mathcal = S \bm a_{\text{in}} \), where \(\bm a_{\text{in} }\)
(\(\bm b_{\text{out}}\)) is a 6-dimensional column vector of incoming
(outgoing) mode amplitudes at each node. In the other valley, the incoming
and outgoing channels of each node are swapped (see Fig.~\ref{fig:network}
(b)).

The matrix elements of \( S \) are determined by microscopic symmetries, such
as \( C_{3z} \) and \( C_{2z} \mathcal T \) symmetry, which preserve the valley
quantum number. Here, \( C_{3z}\) and \( C_{2z} \) represent three-fold and
two-fold rotations about the z-axis through the center of an AA region,
respectively, and \( \mathcal T \) represents time-reversal symmetry.
\begin{align}
	&C_{3z} :  S = C_{3z} S C^{-1}_{3z}, \label{eq:sym1} \\
	&C_{2z}\mathcal T: S = C_{2z}\mathcal T  S (C_{2z}\mathcal T)^{-1} =
	 S^{t}.\label{eq:sym2}
\end{align}
Here, $S^t$ is the transpose of $S$.
These symmetries restrict the scattering matrix to be of the following form:
\begin{align}
	&S = \begin{pmatrix}
		S_1 & S_2 \\
		S_2^{\dagger} & -S_1^{\dagger}
	\end{pmatrix}, \label{eq:smat1} \\
	&S_1 = e^{i \gamma} \sqrt{P_{d1}} \begin{pmatrix}
		0 & 1 & 1 \\
		1 & 0 & 1 \\
		1 & 1 & 0
	\end{pmatrix} 
	+ e^{i \beta} \mathbb I \sqrt{P_{f1}}, \\
	&S_2 = \sqrt{P_{d2}} \begin{pmatrix}
		0 & 1 & -1 \\
		-1 & 0 & 1 \\
		1 & -1 & 0
	\end{pmatrix} 
	- \mathbb I \sqrt{P_{f2}},
\end{align}
where, 
\( S_1 \) and \( S_2 \) incorporate intra-DW-mode and inter-DW-mode
scatterings, respectively. The forward scattering amplitude is given by \( P_f
= P_{f1} + P_{f2} \). Since the scattering matrix \( S \) is unitary, the
remaining amplitude is deflected onto the \( \pm 120^\circ \) rotated domain
walls. The total deflection amplitude is \( P_d = P_{d1} + P_{d2} \). Here, \(
P_{f1} \) and \( P_{f2} \) represent the intra and inter-DW-mode forward
scattering amplitudes, respectively, while \( P_{d1} \) and \( P_{d2} \) denote
the intra and inter-DW-mode deflection amplitudes. The unitarity of \( S \) is
ensured if:
\begin{align}
	&P_f + 2P_d = 1, \\
	&\cos(\beta-\gamma) = \frac{P_{d2}-P_{d1}}{2\sqrt{P_{f1}P_{d1}}} \in [-1,1].
\end{align}
We adopt a symmetric choice of parameters \cite{vakhtel_bloch_2022} as \(
P_{f1} = P_{f_2} = P_{f} / 2 \) and \( P_{d1} = P_{d2} = (1-P_{f}) / 4 \) (we relax this condition
and present some numerical results for \( P_{f1} \neq P_{f2} \) in \cref{sec:asym}). For
this choice, \( \beta = \gamma + \pi / 2 \). This reduces the number of free
parameters in the network to only two: \( P_f \in [0,1] \) and \( \gamma \in
[0,\pi / 2] \). In this phenomenological construction based on symmetry
restrictions, the scattering matrix is assumed to be independent of energy. 

There are two special points in the network-model parameter space with very contrasting
transport properties:
i) at \(P_f = 0\) and \(\gamma = 0\), the system hosts three
independent one-dimensional zig-zag modes, each related by a 120$^\circ$
rotation \cite{Fleischmann_2019,Tsim_2020}, facilitating ballistic transport in MTBG,
and ii) at $P_f = 0$ and $\gamma = \pi / 2$,
where the modes form circulating loops similar to cyclotron orbits in Landau
levels, albeit without an external magnetic field. 
The PLL modes render MTBG insulating \cite{de_beule_aharonov-bohm_2020}.
Keeping \( P_f \) fixed at \( 0 \), by changing \( \gamma \) from 0 to \( \pi / 2 \) one 
interpolates between the zig-zag and pseudo-Landau level modes. 

The position of the nodes is henceforth denoted with a superscript \( (n,m) \), where $n$ is the position index along the junction, and $m$ is the position index in the transverse direction, as illustrated in Fig.~\ref{fig:network} (c).
If one imposes a periodic boundary
condition in the transverse direction, the transverse Bloch momentum becomes a good quantum
number. Incorporating Bloch's theorem, we write the scattering matrix of \(
(n,2) \) nodes in the following way:
\begin{align}
	\label{eq:Sk}
	&\mathcal S^{(n,2)}(k) = U_1(k) \mathcal S^{(n,1)} U_2(k), \\
	&U_1(k) = \mathbb I _{2} \otimes \text{diag}(1,1,e^{-i \sqrt 3 k l}), \\
	&U_2(k) = \mathbb I _{2} \otimes \text{diag}(1,e^{i \sqrt 3 k l},1).
\end{align}
We refer the reader to \cref{Sk derivation} for the derivation of \cref{eq:Sk}.
Here, \( \mathcal S^{(n,1)} = S \), independent of transverse momentum, given by \cref{eq:smat1}.
\( \mathbb I_2 \) is the \( 2\times2 \) identity matrix. 

To construct the Josephson junction, we also incorporate spin  \(\sigma \in \{\uparrow , \downarrow\} \), particle-hole \( s \in \{e
, h\}\) and valley \(\xi \in \{\K,\Kp\}\) indices. Omitting the spin index, we
represent the scattering matrix with the particle-hole and valley index of each
node as a direct sum of the block diagonal matrix in the following way:
\begin{align}
	&\Xi^{(n,1)}(k) = \mathcal S^{(n,1)} \oplus [\mathcal S^{(n,1)}]^* \oplus
	\mathcal S^{(n,1)} \oplus [\mathcal S^{(n,1)}]^*, \\
	&\Xi^{(n,2)}(k) = \mathcal S^{(n,2)}(k) \oplus [\mathcal S^{(n,2)}(-k)]^* \\
	&\hspace{3.5em}\oplus [\mathcal S^{(n,2)}(-k)]^t \oplus [\mathcal
	S^{(n,2)}(k)]^\dagger  \nonumber .
\end{align}
Here, each term in the direct sum ($\oplus$) corresponds to \(\{ (e,\K),
\,(h,\K), \,(e,\Kp), (h,\Kp) \}\) flavour blocks, respectively. The transformation
between these blocks of the net scattering matrix can be obtained by the operations of
time reversal (\(\mathcal T\)) and charge conjugation (\( \mathcal C \)), as summarized in Table~\ref{tab:transform}. For more details about the action of symmetry operators on the scattering matrix, we refer the reader to the Appendix~\ref{sec:app:symm}.

We construct the full
scattering matrix \( S_{\text{node}} \) of the network with the scattering
matrices of all the nodes \cite{baxevanis_even-odd_2015},
\begin{align}
	S_{\text{node}}(k) = \sigma_0 \otimes \bigoplus_{n,m} \Xi^{(n,m)}(k).
\end{align}
Here, \( \sigma_0 \) is the \( 2\times 2 \) identity matrix in the spin-space
and reflects the fact that the intrinsic system retains spin rotation symmetry.
\( S_{\text{node}}(k) \) is a sparse matrix, and for the geometry that we
consider for the JJ as shown in Fig.~\ref{fig:network} (c), the dimension of
the matrix is \( \mathcal D = (2N-1) \times2^3 \times 6 \), where \( 2N - 1 \)
is the total number of nodes present in the network. The \( S_{\text{node}} (k)
\) matrix acts on the incoming modes of the whole network and returns the
outgoing modes of the entire network, represented by 
\begin{align*}
\bm{b} = S_{\text{node}}(k)\, \bm{a},
\end{align*}
here \( \bm{a} \) and \( \bm{b} \) are column
vectors with dimensions \( \mathcal D \) that represent the incoming and
outgoing modes of the entire network, respectively.
\begin{table}[htpb]
	\caption{ 
 Summary of symmetry operators (denoted by $\mathcal O$ in the fifth column) and their action on the scattering matrix (in the last column). The first column denotes the node index. The columns from second to fourth represent the particle/hole, valley, and spin flavors, respectively. We derive the scattering matrix for each flavor by using the symmetry operation $\mathcal O$. Since there is no spin-orbit coupling in the system, the symmetry operator's action is identical for the $\uparrow$ and $\downarrow$ spins. Note that the momentum enters through the $\mathcal S^{(n,2)}$ matrices only. For more details about the action of symmetry operators on the scattering matrix, we refer the reader to Appendix~\ref{sec:app:symm}.
 }
	\begin{ruledtabular}
		\begin{tabular}{cccccl}
			Node & PH  & Valley & Spin  &  $\mathcal O$ & \(  \mathcal O \mathbb S \mathcal O^{-1} \) \\[5pt]
			\hline
			\((n,1)\) &\(e\) & \(\K\) & \( \uparrow / \downarrow \) & \( \mathbb
			I \) & \( \mathcal S^{(n,1)} \equiv \mathbb S\) for $(n,1)$ nodes \\
			\((n,1)\) &\(h\) & \(\K\) & \( \uparrow / \downarrow \) & \( \mathcal
			C \) & \( [\mathcal S^{(n,1)}]^* \) \\
			\((n,1)\) &\(e\) & \(\Kp\) & \( \uparrow / \downarrow \)& \( \mathcal
			T \) & \( \mathcal [S^{(n,1)}]^t = S^{(n,1)} \) (Eq.~\eqref{eq:sym2})\\
			\((n,1)\) &\(h\) & \(\Kp\) & \( \uparrow / \downarrow \)& \( \mathcal
			T \mathcal C \) & \( [\mathcal
			S^{(n,1)}]^\dagger = [S^{(n,1)}]^* \) (Eq.~\eqref{eq:sym2}) \\
			\hline
			\((n,2)\) &\(e\) & \(\K\) & \( \uparrow / \downarrow \) & \( \mathbb
			I \) & \( \mathcal S^{(n,2)}(k) \equiv \mathbb S \) for $(n,2)$ nodes  \\
			\((n,2)\) &\(h\) & \(\K\) & \( \uparrow / \downarrow \) & \( \mathcal
			C \) & \( [\mathcal S^{(n,2)}(-k)]^* \) \\
			\((n,2)\) &\(e\) & \(\Kp\) & \( \uparrow / \downarrow \)& \( \mathcal
			T \) & \( [\mathcal S^{(n,2)}(-k)]^t \) \\
			\((n,2)\) &\(h\) & \(\Kp\) & \( \uparrow / \downarrow \)& \( \mathcal
			T \mathcal C \) & \( [\mathcal S^{(n,2)}(k)]^\dagger \) \\
		\end{tabular}
	\end{ruledtabular}
	\label{tab:transform}
\end{table}

Following each scattering event, the outgoing modes propagate freely for a
duration of \( \tau = l / v_F \) before the subsequent scattering event. During
this time, the DW modes acquire a dynamic phase \( \xi \epsilon l/ \hbar v_F \), where \(
\epsilon \) is the energy of the propagating modes. The valley index in the
dynamical phase accounts for the DW modes propagating in opposite
directions in different valleys. A
recent STM study \cite{Huang_2018} has shown that the DW modes are physically separated
by a large length scale compared to the atomic scale. Hence, we assume that the DW modes are decoupled along the links of the network. After time $\tau$, the outgoing and incoming modes between the neighboring nodes are given by
\begin{align}
    \bm a^{(n,m;\eta)}_{i\xi} = e^{i \xi \epsilon l / v_F} \bm b^{(\overline n, \overline m;\eta)}_{i\xi},
    \label{eq:link1}
\end{align}
where $(\overline n, \overline m)$ is a neighboring node of $(m,n)$, $\bm a_{i\xi}^{(n,m)}$ and $\bm b_{i\xi}^{(\overline n,\overline m)}$ represent the amplitudes of incoming and outgoing modes for these nodes, respectively. The modes are \emph{four-dimensional} vectors in the basis of $\{e,h\} \otimes \{\uparrow,\downarrow\}$ for each of the DW modes. Here, $i \in \{1,2,3\}$ refers to the three directions of the channels, and $\xi \in \{\K,\Kp\}$ denotes the valley index, \( \eta \in \{1,2\} \) runs
 over two DW modes per valley, per spin.

At the superconducting electrodes, an electron of \( \K \) valley and \( \uparrow \)-spin is Andreev reflected as a hole of \( \Kp \) valley with \( \downarrow \)-spin and vice versa. 
As the links are made from one-dimensional DW modes, only retro Andreev reflection takes place,
and specular Andreev reflection is suppressed \cite{beenakker_specular_2006}.
The Andreev reflection is incorporated in the left and right lead via the following matrix \cite{baxevanis_even-odd_2015,Beenakker_2013}:
\begin{align}
	M_{A}(\phi) = i \alpha
	\begin{pmatrix}
		0 & 0 & 0 & -e^{i \phi} \\
		0 & 0 & e^{i \phi} & 0 \\
		0 & e^{-i \phi} & 0  & 0 \\
		-e^{-i \phi} & 0 & 0 & 0
	\end{pmatrix}.
\end{align}
\( \alpha \) is defined as,
\begin{align}
	\alpha = 
	\begin{cases}
		i \exp(- i \cos^{-1}\left({\epsilon}/{\Delta}\right))\quad \text{for } \epsilon \le  \Delta \\
		i \exp(- \cosh^{-1}\left({\epsilon}/{\Delta}\right)) \quad \text{for } \epsilon > \Delta,
	\end{cases}
\end{align}
where $\Delta$ and $\phi$ are the superconducting gap and phase, respectively. The Andreev reflection connects outgoing and incoming modes on the left and right
superconducting junction as follows:
\begin{align}
    \bm a^{(n,m;\eta)}_{i \K} &= M_A(\phi / 2) \bm b^{(n,m;\eta)}_{i \K'},~\text{$(m,n) \in $ left junction}\label{eq:sc-junction1}\\
    \bm a^{(n,m;\eta)}_{i \K'} &= M_A(-\phi / 2) \bm b^{(n,m;\eta)}_{i \K},~\text{$(m,n) \in $ right junction}.\label{eq:sc-junction2}
\end{align}

The equation \eqref{eq:link1} for the MTBG links
and \eqref{eq:sc-junction1}--\eqref{eq:sc-junction2} for the
left and right superconducting leads and normal MTBG junction define the bond matrix \(S_{\text{bond}} \) that acts on outgoing modes and returns incoming modes,
i.e., 
\begin{align*}
\bm{a} = S_{\text{bond}}(\epsilon,\phi) \bm{b}.
\end{align*}
 For details of the $S_{\text{bond}}(\epsilon,\phi)$ matrix we refer the reader to the Appendix~\ref{sec:app:sbond}.
\begin{figure*}[ht]
	\centering
	\includegraphics[width=510pt]{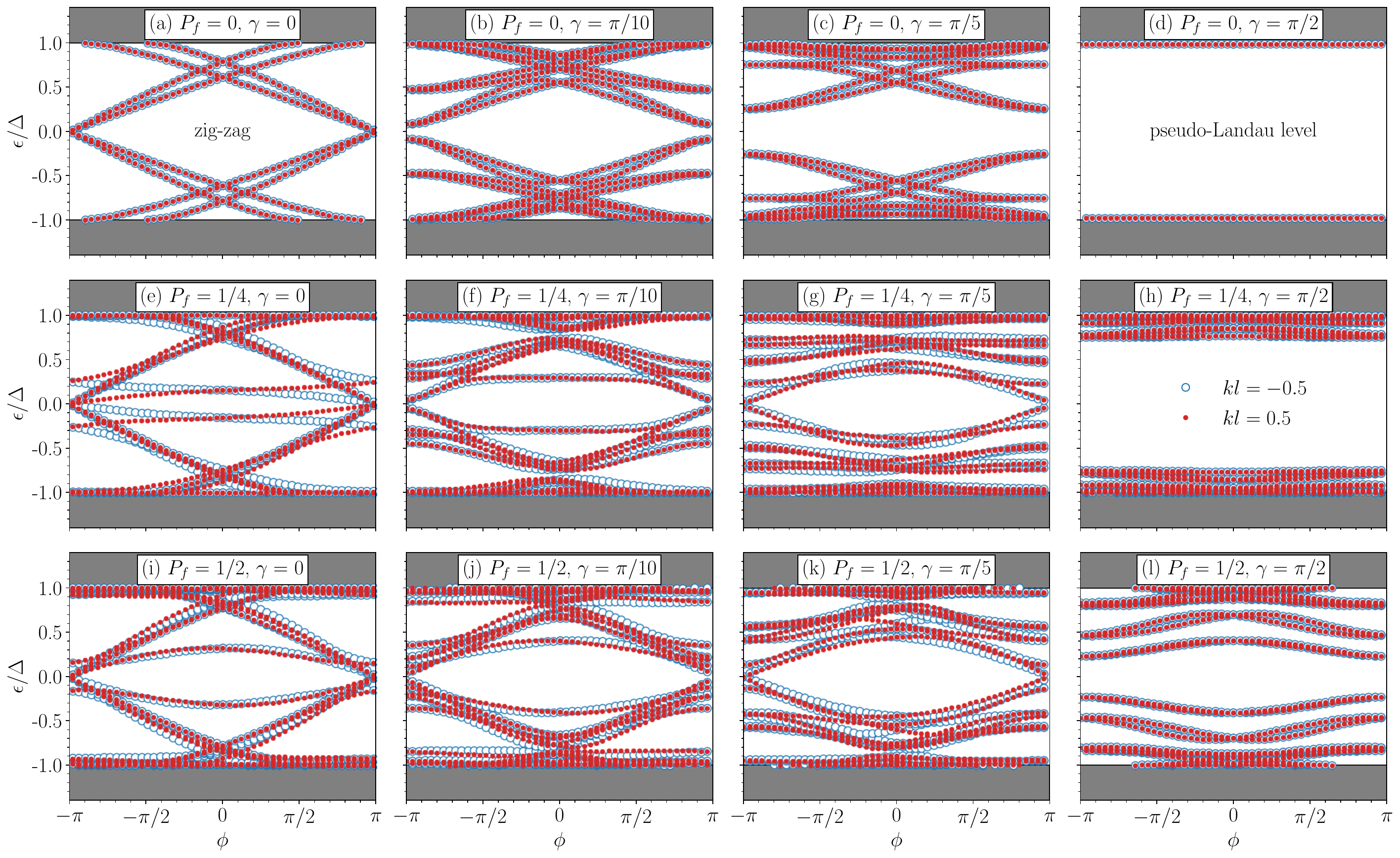}
	\caption{Andreev bound states (ABS) with different choice of \( P_f \) and
		\( \gamma\) and for two representative values of transverse momentum \(
		k l = 0.5 \) (red) and \(k l = -0.5\) (blue). The gray regions denote the continuum of states \( \abs{\epsilon} >
		\Delta \). For (a) \( P_f = 0,\gamma = 0 \), we have the zig-zag modes
		in the network which has linearly dispersing ABS near zero energy. (a)
		is the ABS for zig-zag modes hosting zero energy states at \(\phi = \pm \pi\). The zero-energy ABS persists if we increase $P_f$, while keeping $\gamma$ fixed at zero, see (a)$\to$(e)$\to$(i). As we increase \( \gamma \) keeping \(
		P_f \) constant, the ABS becomes gapped, and the bandwidth decreases, and 
  at \(P_f = 0, \gamma = \pi / 2 \), it becomes perfectly flat for the pseudo-Landau
		level modes. We use $\theta=0.1^\circ$, $N=3$, $\Delta=1$~meV for these numerical plots.
  }
	\label{fig:abs}
\end{figure*}
\begin{figure*}[ht]
	\centering
	\includegraphics[width=\linewidth]{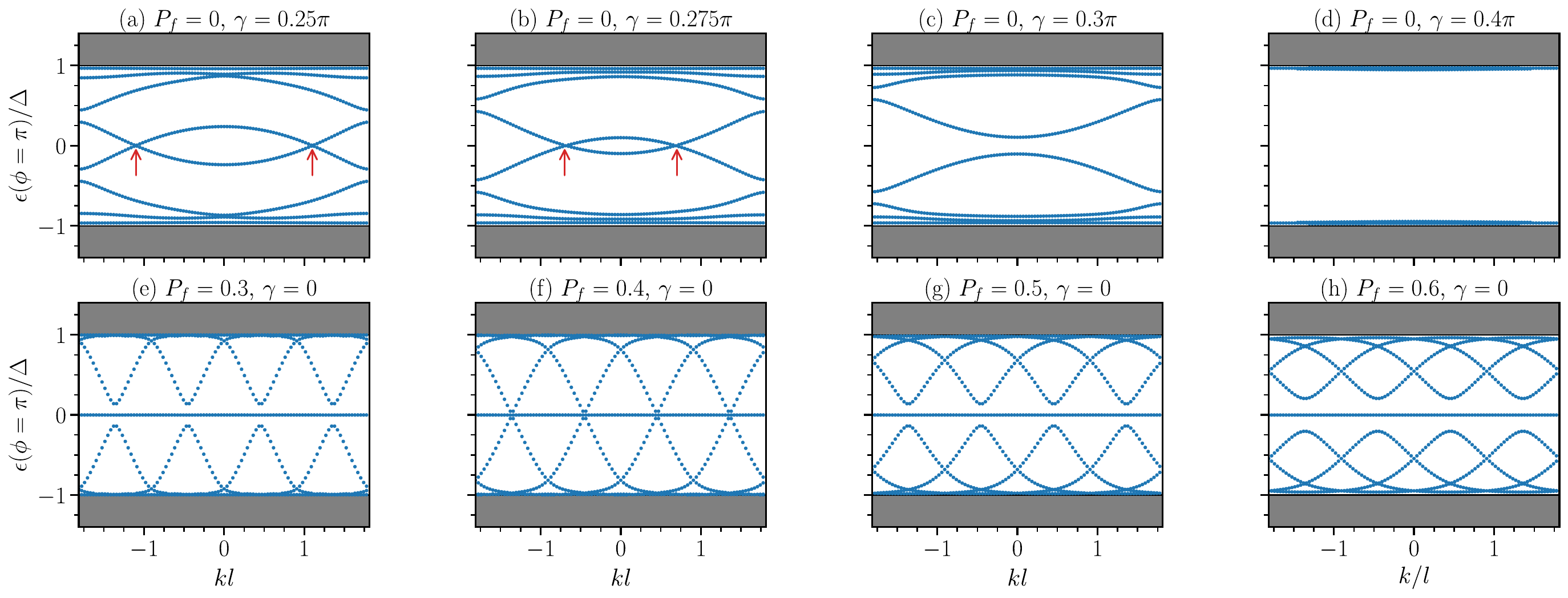}
	\caption{Andreev bound states (ABS) for \( \phi = \pi  \) over the first
		Brillouin zone, \( k l \in [  -2\pi / \sqrt 3, 2\pi / \sqrt 3)
		\), for several values of \( (P_f, \gamma) \). (a)-(d) shows evolution of
		the ABS as we change \( \gamma \) keeping \( P_f = 0 \), the system evolves from a
		gapless-ABS spectrum to gapped-ABS state. The two Dirac-cones (indicated by red arrows) in the \( k-\phi \) plane merges and
		gap out.
		(e)-(h) shows the evolution of ABS bands when we change \( P_f \in [0.3,0.6] \) 
		keeping \( \gamma=0 \) fixed. The ABS always have a momentum-independent 
		zero energy mode at \( \phi = \pi \) which disperses linearly in the \(
		\phi \) direction (See Fig.~\ref{fig:abs}). There is a gap-closing that happens at
		\( P_f \approx 0.4 \). 
  For these numerical plots, we use $\theta=0.1^\circ$, $N=3$, and $\Delta=1$~meV.
}
	\label{fig:k-phi}
\end{figure*}
\section{Andreev bound states and Josephson current}
\label{sec:abs}
Modes with energies smaller than the superconducting gap (\(\abs{\epsilon} \le {\Delta}\)) can not propagate through the superconductor. Consequently, they undergo multiple Andreev reflections at the superconducting interfaces, leading to the formation of bound states known as Andreev bound states (ABS).
To find the ABS, we first note that \( S_{\text{node}}(k) \)
and \( S_{\text{bond}}(\epsilon,\phi) \) matrices satisfy the equation \( (\mathbb{I}
-S_{\text{node}}(k)S_{\text{bond}}(\epsilon,\phi))\bm{a} = 0 \). Then, for a non-trivial solution of
\( \bm{a} \), the following condition must hold:
\begin{align}
	\label{eq:abs-trans}
	\det(\mathbb{I} -S_{\text{node}}(k)S_{\text{bond}}(\epsilon,\phi) ) = 0.
\end{align}
The above determinantal equation is a transcendental equation that needs to be solved to find
the energy of ABS (\( \epsilon \)) as a function of the superconducting phase
difference \( \phi \) and Bloch momentum \( k \). The discrete ABS of the
junction are denoted by \( {\epsilon_p(\phi, k)} \), where \( p \) is the band index for ABS.

Each of the ABS at zero temperature (\( T \)) contributes to the Josephson current
by an amount \( J_p(\phi) = \frac{2e}{\hbar}\int \frac{\dd k}{ 2\pi}
\dv{\epsilon_p(\phi,k)}{\phi} \). Here, we adopt a more general framework \cite{PhysRevLett.67.3836, BROUWER19971249, Beenakker_2013, baxevanis_even-odd_2015,PhysRevLett.68.1255} to calculate the Josephson current at a finite temperature that takes into account the contribution coming from the quasiparticle continuum into the Josephson current as well. The expression of Josephson current reads,
\begin{align}
	J(\phi) &= -k_B T \frac{2 e}{\hbar}\frac{\sqrt 3 l}{2\pi} \int_0^{2\pi / \sqrt 3 l} 
	 \dd k   \nonumber\\
	&\hspace{2em}\times \dv{}{\phi}\sum_{p = 0}^{\infty}\ln \det[1 -
	S_{\text{node}}(k) S_{\text{bond}}(i\Omega_p,\phi)],
	\label{eq:current}
\end{align}
where the sum is over the fermionic Matsubara frequencies \( \Omega_p = (2p +
1) \pi k_B T \). Using Eq.~\eqref{eq:current}, we compute the Josephson current
numerically for different network parameters of MTBG. For \( T \to 0 \),
the Matsubara summation becomes an integration, i.e., \( k_B T \sum_{p} \to \int
\frac{\dd \omega}{2 \pi} \). Eq.~\eqref{eq:current} is valid under the
assumption that the system reaches equilibrium without restrictions on the
fermion parity; therefore, it holds for time scales that are much longer
compared to the quasiparticle poisoning time \cite{Beenakker_2013}. This is the approximation we
adopt in this work to compute the Josephson current.
If this is not the case, corrections for parity conservation will be necessary \cite{baxevanis_even-odd_2015}.

Due to the large moiré length scale ($l$) for smaller $\theta$, the MTBG-Josephson junction naturally falls into the category of a \emph{long junction}, where the junction length $L=N l$ greatly exceeds the coherence length of the superconductor $\xi$  ($L \gg \xi$). The opposite limit ($L \ll \xi$) is dubbed as a \emph{short junction}.
\section{Numerical results}
\label{sec:numerics}
\begin{figure*}[ht]
	\centering
	\includegraphics[width=510pt]{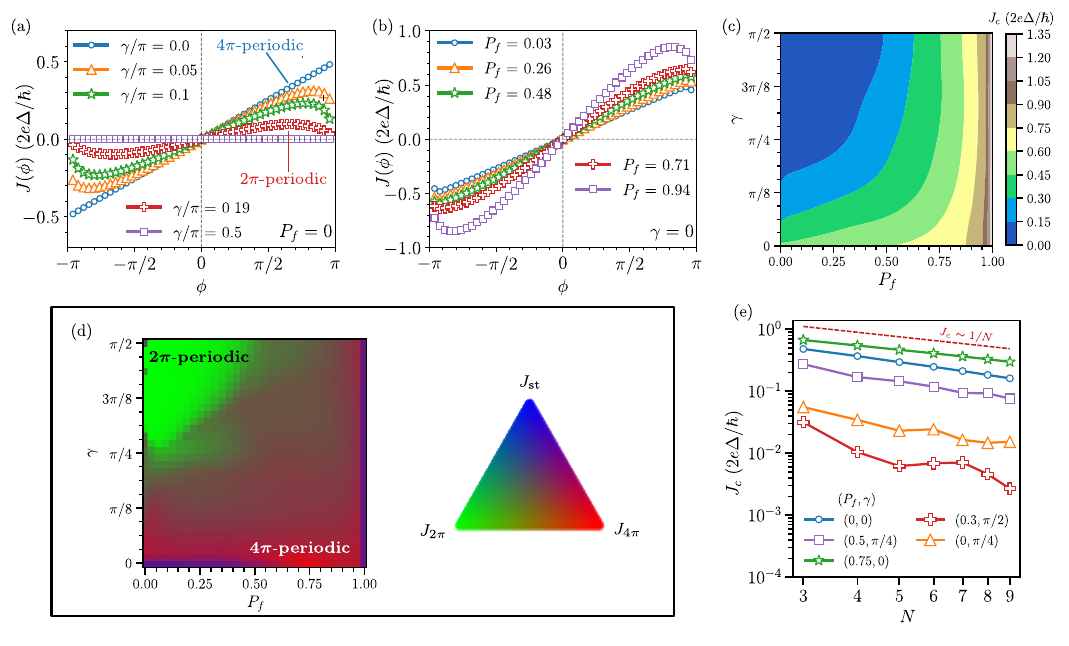}
	\caption{(Color online)
 (a) Josephson current for \(P_f = 0\), and several values of \(\gamma\). As we change \(\gamma\) to go from the zig-zag modes to pseudo-Landau level modes, the Josephson current transforms from a \(4\pi\) periodic sawtooth current profile to a \(2\pi\) periodic sinusoidal current profile. For pseudo-Landau level modes ($P_f = 0, \gamma = \pi / 2$) the Josephson current becomes zero. 
 (b) Josephson current for \(\gamma = 0\), and several values of \(P_f\). By tuning \(P_f\), the Josephson current changes from a sawtooth to a sinusoidal 4$\pi$-periodic profile while maintaining the \(4\pi\) periodicity.
 (c) Critical current for all values of \((P_f,\gamma)\). The critical current is zero for the pseudo-Landau level modes ($P_f = 0, \gamma = \pi / 2$) and increases monotonically as we move away from this parameter by decreasing \(\gamma\) and increasing \(P_f\).
 (d) The Josephson current profile for all values of \((P_f, \gamma)\). In the Josephson junction of MTBG, the current profile resembles one or a combination of the following: i) \(2\pi\) periodic sinusoidal current (\(J_{2\pi}\)), ii) \(4\pi\) periodic sinusoidal profile (\(J_{4\pi}\)) and iii) \(4\pi\) periodic sawtooth profile (\(J_{\mathrm{st}}\)) represented by green, red and blue colors, respectively. The mixed character of \(J_{\text{st}/2\pi/4\pi}\) is represented by the appropriate blending of red, green, and blue (see \cref{eq:blend}). (e) The critical current dependence on the length ($Nl$) of the Josephson junction for representative network parameters that show a \(4\pi\) and \(2 \pi\) periodic Josephson current profile.
 For these numerical plots we use $\theta=0.1^\circ$, $\Delta=1$~meV and $T \ll \Delta = 0.01 \Delta$. For (a)--(d), we use $N=3$.
}%
	\label{fig:Josephson current}
\end{figure*}
\subsection{Andreev bound-state (ABS) spectrum}
In Fig.~\ref{fig:abs}, ABS is illustrated for several values of the network
parameters and two representative values of Bloch momentum \( kl =
\{-0.5,0.5\} \). The time reversal and particle-hole symmetry present in the system
imposes that \( \epsilon_{p}(\phi,k) = \epsilon_{p}(-\phi,-k)\) and \(
\epsilon_p(\phi,k) = -\epsilon_{-p}(\phi,-k) \), respectively.  Moreover, the ABS spectrum is independent of Bloch momentum in the presence of PLL modes, i.e., \(
P_f = 0 \) and \( \gamma = \pi / 2 \) (see \cref{fig:abs} (d)).
If we keep $\gamma = 0$ and increase the value of
$P_f$ (see Fig.~\ref{fig:abs} (a), (e), (i)), we find that the zero-energy crossing at \( \phi = \pm \pi \) is robust under this change. 
On the contrary, with increasing \( \gamma \), the zero energy crossing at \( \phi = \pm \pi \) becomes fully gapped.

For the ZZ modes, i.e., when $P_f = 0$ and $\gamma = 0$, the ABS near zero energy varies 
linearly as a function of the superconducting phase
difference, i.e., $\epsilon \sim (\pi - \abs{\phi})$, featuring a zero energy
crossing at \( \phi = \pm \pi \). As the ABS merge into the quasiparticle continuum, it becomes curved (see Fig.~\ref{fig:abs} (a)). Such ABS have been previously reported for a Josephson junction on the edge of a quantum spin Hall insulator \cite{Beenakker_2013}. 
With increasing \( \gamma \) while keeping \( P_f \) fixed at zero, the ABS are no
longer linear in the phase difference. 
At the PLL parameter regime, i.e., when \( P_f = 0 \) and \( \gamma = \pi / 2 \), the MTBG hosts circulating modes and all the
ABS above and below zero energy coalesce into two perfectly flat bands (see Fig.~\ref{fig:abs} (d)). 

As we move away from $P_f=0$ and $\gamma=0$ line in the parameter space, many more ABS appear. 
For several network parameters (e.g., see \cref{fig:abs} (f), (g), (j), (k) and also \cref{fig:abs-ex} in \cref{app:Extended data for Andreev bound states}), we see that there coexist zero-energy gapless and fully gapped ABS.
As we discuss in the following section, this coexistence results in a skewed Josephson current.

We also investigate the ABS with respect to Bloch momentum in
Fig.~\ref{fig:k-phi}. We particularly focus on \( \phi = \pi \), where the presence of zero-energy ABS is plausible.
 For \( P_f = 0 \) and for finite \( \gamma \) (see
Fig.~\ref{fig:k-phi} top panel), the ABS hosts zero energy states and forms two
Dirac cones in the \( k\)--\(\phi \) space. As we increase \( \gamma \) while
keeping \( P_f \) fixed at zero, we see that the two Dirac cones merge at a certain value
of \( \gamma \) and gap out, further increasing \( \gamma \) makes the ABS
flatter as we approach \( \gamma = \pi / 2 \). This is how the ABS spectrum
becomes gapped as one approaches the PLL limit. 
This mechanism leads to a transition from a \( 4\pi \) periodic Josephson current
to a \( 2\pi \) periodic one.
Similarly, if we fix \( \gamma  \) to 0 and increase \( P_f \) (see
Fig.~\ref{fig:k-phi} bottom panel), the ABS always exhibits a zero energy state
for all values of \( P_f \). This is the zero energy ABS, which has a linear
dependence on \( \phi \) as mentioned previously (see Fig.~\ref{fig:abs} (a)). Upon increasing \( P_f \), at
a certain value of \(P_f\), the gap between the zero-energy ABS and finite
energy bands closes. However, this gap reappears as \( P_f \) is
increased further. During this ABS gap closing, the Josephson current changes from
a sawtooth to a sinusoidal 4$\pi$-periodic profile, as discussed below.

\subsection{Josephson current}
For the MTBG network, the Josephson responses have been summarized in Fig.~\ref{fig:Josephson current}, where we observe that the form of Josephson current depends strongly on network parameters. We note that the zero-energy level crossings in ABS induce a $4\pi$ periodicity of the Josephson current. On the other hand, gapped ABS always results in a $2\pi$ periodic Josephson current. As shown in \cref{fig:abs}, a combination of gapped and gapless ABS may also coexist for several network parameters. 
In such cases, Josephson current exhibits a mixed nature, comprising $2\pi$ and $4\pi$ periodic components, resulting in a skewed current phase relationship.

 As illustrated in Fig.~\ref{fig:Josephson current} (a) in the zig-zag limit ($P_f = 0, \gamma = 0$), a
sawtooth Josephson current is observed, i.e.,
\begin{align}
  J_{\text{st}}(\phi) \sim \frac{2 e\Delta }{\hbar}
\phi,\,\abs{\phi} < \pi,
\label{eq:J-st}
\end{align}
with a discontinuity at \( \phi = \pm \pi \). Such discontinuity in the Josephson current is a
signature of \( 4\pi \) periodicity. This Josephson current profile resembles that of a long normal metallic Josephson junction \cite{ishii1970josephson}, and a long Josephson junction at the edge of a quantum spin Hall insulator \cite{Beenakker_2013}. 
This discontinuity arises from the presence of zero-energy Andreev bound states
(ABS), as depicted in \cref{fig:abs} (a) and \cref{fig:k-phi} (a) and due to the assumption that the system equilibrates without any parity constraint. In
this regime, perfect ballistic transmission is facilitated by the zig-zag modes,
even when the forward scattering ($P_f$) is zero. Keeping $P_f$ fixed at zero, as we increase \(
\gamma \), the Josephson current profile changes from sawtooth ($J_{\text{st}}$) to a
sinusoidal \( 2\pi \) periodic current, resembling a traditional Josephson
current phase relation: 
\begin{align}
  J_{2\pi}(\phi) \sim \frac{2 e\Delta }{\hbar} \sin(\phi).
\label{eq:J-sine}
\end{align}
The $2\pi$ periodic Josephson current indicates that the ABS spectrum is fully gapped near zero-energy. The transition from $J_{\text{st}}$ to $J_{2\pi}$ occurs when two Dirac cones of ABS in the $k-\phi$ space merge and become fully gapped (see \cref{fig:k-phi} top panel). Further increasing \( \gamma \) leads to a decrease in the amplitude of the Josephson current, reaching zero finally at \( \gamma = \pi / 2 \). This phenomenon can be attributed to the circulating modes for the PLL network parameters, which do not support transport through the bulk, as evidenced by the perfectly flat ABS (see \cref{fig:abs} (d)).

On the other hand if we keep \( \gamma \) fixed at 0, and increase \( P_f \),
the Josephson current profile transforms from a sawtooth to a sinusoidal $4\pi$-periodic Josephson current:
\begin{align}
  J_{4\pi}(\phi) \sim \frac{2 e\Delta }{\hbar} \sin(\phi/2),\,\abs{\phi} < \pi.
\label{eq:J-4pi}
\end{align}
There are zero-energy Andreev bound states present in the spectrum that facilitate this $J_{4\pi}$ periodic Josephson current. The transformation from $J_{\text{st}}$ to $J_{4\pi}$ happens at $P_f \approx 0.4$, which coincides with the value of \( P_f \) where the gap between the zero-energy and finite energy ABS vanishes (see Fig.~\ref{fig:k-phi} bottom panel). 

We study the dependence of critical current \( J_c \,(= \max_\phi \abs{J(\phi)})\) in
Fig.~\ref{fig:Josephson current} (b), on the MTBG network parameters \( P_f \) and \( \gamma \). As
mentioned previously, the critical current vanishes for PLL modes, i.e., \( P_f = 0 \) and \(\gamma = \pi / 2 \) and increases monotonically as we move away from that
point by decreasing \(\gamma\) and increasing \(P_f\). 

As one deviates from the $P_f=0$ and $\gamma=0$ lines in the parameter space, the Josephson current exhibits a combination of $J_{\text{st}/2\pi/4\pi}$ responses. We devise a method to determine which of the three responses or combinations thereof the Josephson current profile closely resembles. 
First, we normalize \( J( \phi) \) so that it has the same amplitude as $J_{\text{st}/2\pi/4\pi}$. Then, the distances between the
functions \( J(\phi) \) and $J_{\text{st}/2\pi/4\pi}$ is calculated. 
This distance serves as a criterion to determine which current profile \(
J(\phi) \) resembles closely. The distance between two functions is
formally defined as \( ||f - g|| = \sqrt{\braket{f-g}{f-g}} \), with the inner product
defined as \( \braket{f}{g} = \int_0^{2\pi} \, \dd \phi f(\phi) g(\phi) / 2 \pi \).
We construct a 3-tuple from the inverse distances as
\begin{align}
	&\{u_1, u_2, u_3\} \propto
 \left\{\frac{1}{||J - J_{\text st}||},\frac{1}{||J -
	J_{2 \pi}||}, \frac{1}{||J - J_{4 \pi}||}\right\},
 \label{eq:blend}
\end{align}
which is normalized such that $\sum_i u_i = 1$.
The 3-tuple $\{ u_1,  u_2, u_3\}$ represents the barycentric coordinates of an equilateral triangle. The coordinates: $\{1,0,0\}$, $\{0,1,0\}$,
$\{0,0,1\}$ refer to the Josephson current profile of \( \{J_{\text{st}},
J_{2\pi}, J_{4\pi}\} \), respectively. We assign the colors blue,
green and red to these coordinates, respectively. For any other coordinates,
we blend the colors (blue, green, red) in the ratio $\{ u_1,  u_2,
 u_3\}$.  In this manner, we can
discern which profile closely resembles \( J(\phi) \). 

The classification map of
Josephson current is presented in Fig.~\ref{fig:Josephson current} (d), across the
entire parameter space of \((P_f, \gamma)\). This map correctly captures the transition from $J_{\text{st}}$ to $J_{4\pi}$ at $P_f \approx 0.4$ and $\gamma = 0$. Around the PLL parameters is a region where the Josephson current resembles $J_{2\pi}$. As \( P_f \) reaches 1, irrespective of the values of \( \gamma \in [0,\pi / 2] \), the Josephson current profile becomes $J_{\text{st}}$ again. For most other regions of the
parameter space, a mixed \( J_{2\pi} \) and \( J_{4\pi} \) character is observed. \cref{fig:Josephson current} (c), we also note From regions with the $J_{2\pi}$ Josephson current has a smaller critical current than that of $J_{\text{st}}$ and $J_{4\pi}$ regions.

In Fig.~\ref{fig:Josephson current} (e), we show the dependence of the critical
current as a function of the network length \( N l \), for various network
parameters. 
For parameters resulting in a $4\pi$-Josephson current, we observe that the critical current decays as \( \sim 1 / (Nl) \). This behavior has been demonstrated for a long Josephson junction with normal metal barriers \cite{ishii1970josephson} and Josephson junction at the edge of a quantum spin Hall insulator \cite{Beenakker_2013}.
 For some representative values
of parameters, this is shown in Fig.~\ref{fig:Josephson current} (e). On the other
hand, if the current is \( 2 \pi\) periodic, the critical current decreases in
a non-monotonic manner.

\begin{figure*}[ht]
	\centering
	\includegraphics[width=\linewidth]{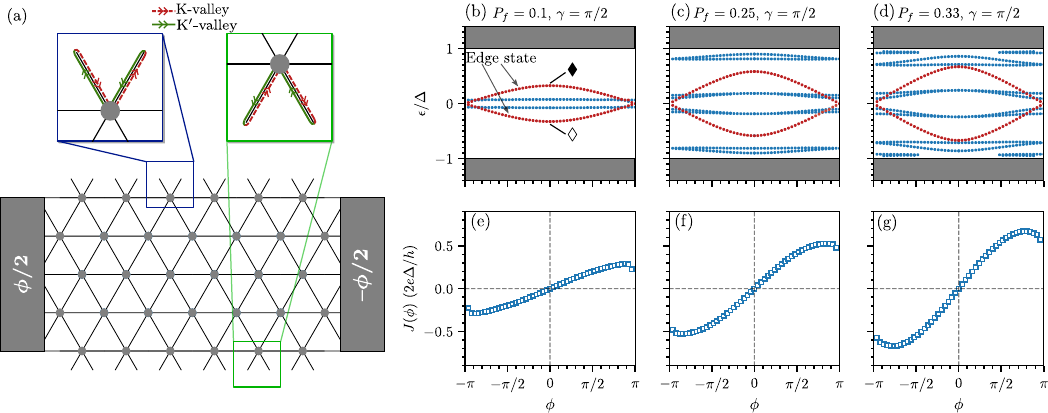}
	\caption{
(a) The Josephson junction of a minimally twisted bilayer graphene with open top and bottom edges. At the edges, modes in K-valley are reflected back to K$'$-valley and vice versa (see \cref{eq:edge1,eq:edge2,eq:edge3,eq:edge4}). (b)--(d) Andreev bound states (ABS) and corresponding Josephson current (e)--(g) in the presence of open edges in the network. Then ABS, which arises in the presence of the edges, are annotated and shown in red.
In the presence of pseudo-Landau level modes, the edges mediate the Josephson current, even if it is zero in the bulk.
For the plots we use $\theta=0.1^\circ$, $(N_x,N_y)=(3,5)$, $\Delta = 1$~meV and $T=0.01\Delta$.	}
	\label{fig:edge-newtork}
\end{figure*}
\section{Effect of edges}
\label{sec:edge}
So far, we imposed periodic boundary
conditions in the transverse direction. In this section, we relax this condition
and examine the impact of physical edges on ABS and Josephson current. In general, the geometric form of edges can be complicated because the relaxation at the boundaries of the MTBG can be
different from that of the bulk. In order to understand the quantitative structure of
the edges, extensive first-principle studies may be required, which, to the best of our
knowledge, have not yet been documented in the literature. Following
\cite{Chou_2020}, we use a theoretical model of the edges and assume that the
domain-wall modes of K-valley can be perfectly reflected back to K$'$-valley up
to a phase and vice versa at the edges (see Fig.~\ref{fig:edge-newtork} (a)). Such scatterings at the
edges keep the network's time reversal symmetry intact. This model
additionally assumes that the truncation of the edges is located far away from
the scattering nodes (i.e., AA regions) of the network. This is to ensure that the
microscopic symmetries (Eq.~\eqref{eq:sym1} and \eqref{eq:sym2}) of the
scattering matrices near the edges are not destroyed. To this extent, we
consider a finite network of size \( (N_x, N_y) \), as shown in
Fig.~\ref{fig:edge-newtork} (a). 

The scattering matrix reads as follows:
\begin{align}
	&\Xi^{(n,m)} = \mathcal S^{(n,m)} \oplus [\mathcal S^{(n,m)}]^* \oplus
	\mathcal S^{(n,m)} \oplus [\mathcal S^{(n,m)}]^*, \\
	&S_{\text{node}} = \sigma_0 \otimes \bigoplus_{n,m}\Xi^{(n,m)}.
\end{align}
The boundary condition for the domain-wall modes for the edges reads as follows:
For the top edge, we have:
\begin{align}
	\bm a_{2\K}^{(n,N_y;\eta)} &= \mathrm{diag}(e^{-i \psi_1 }, e^{i \psi_1 }, e^{-i \psi_1 }, e^{i \psi_1 }) \bm b_{2\Kp}^{(n,N_y;\eta)} \label{eq:edge1}, \\
	 \bm a_{3\Kp}^{(n,N_y;\eta)} &= \mathrm{diag}(e^{i \psi_1 }, e^{-i \psi_1 }, e^{i \psi_1 }, e^{-i \psi_1 }) \bm b_{3\K}^{(n,N_y;\eta)}.
  \label{eq:edge2}
\end{align}
Similarly, for the bottom edge, we have:
\begin{align}
	\bm a_{2\Kp}^{(n,1;\eta)} &= \mathrm{diag}(e^{-i \psi_2 }, e^{i \psi_2 }, e^{-i \psi_2 }, e^{i \psi_2 }) \bm b_{2\K}^{(n,1;\eta)}, 
 \label{eq:edge3}
 \\
	 \bm a_{3\K}^{(n,1;\eta)} &= \mathrm{diag}(e^{i \psi_2 }, e^{-i \psi_2 }, e^{i \psi_2 }, e^{-i \psi_2 }) \bm b_{3\Kp}^{(n,1;\eta)}.
\label{eq:edge4}
\end{align}
 Here, \( i \in \{1,2,3\} \) in the subscript of the modes \(\{i \xi\}\) refers
 to the three incoming and outgoing channels, \(\xi \in \{\K,\Kp\}\) is the valley index, \( \eta \in \{1,2\} \) runs
 over two DW modes per valley, per spin. Here, \( \bm a_{i\xi}^{(n,m;\eta)} \) and  \( \bm b_{i\xi}^{(n,m;\eta)} \) represent the amplitudes of the incoming and outgoing modes by \emph{four-dimensional} vectors written in the basis of  \( \{e,h\} \otimes \{\uparrow,\downarrow\}\). \( \psi_1 \) and \( \psi_2 \) are the parameters associated with the reflection at the top and bottom edges, respectively. The ABS
and Josephson current depends weakly on these parameters, and without loss of generality,
we choose \( \psi_1 = \psi_2 = 0 \) for our calculations.

In addition to Eq.~\eqref{eq:link1}, which describes the link connections in the MTBG, and Eq.~\eqref{eq:sc-junction1}--\eqref{eq:sc-junction2} which describe the connections between the MTBG network and superconducting leads, equations \eqref{eq:edge1}--\eqref{eq:edge2} are the newly introduced equations for the top and bottom edges that are included in the \( S_{\mathrm{bond}} \) matrix. Note that the superconducting leads connect the K and K$'$ valleys through Andreev reflection, whereas the top and bottom edges of the network connect the two valleys via normal reflection.

We focus on the pseudo-Landau level regime because, in this parameter range, the bulk does not allow for any Josephson current. Instead, the Josephson current is mediated through the edges of the network created by the domain-wall modes.
\begin{figure}
    \centering
    \includegraphics[width=246pt]{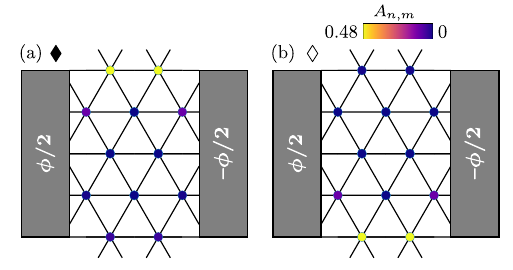}
    \caption{Amplitudes ($A_{n,m}$) of the incoming modes for the dispersive ABS shown in Fig.~\ref{fig:edge-newtork} (b). 
    (a) shows the amplitudes for ABS marked by $\blacklozenge$ in Fig.~\ref{fig:edge-newtork} (b). The amplitude is localized on the top edge for this ABS. (b) shows the same for ABS marked by $\lozenge$. The amplitude is localized on the bottom edge. Here we use a network of size $(N_x,N_y)=(3,5)$.}
    \label{fig:edge-state}
\end{figure}
We find the ABS by solving \cref{eq:abs-trans} with the modified \(
S_{\mathrm{node}} \) and  \( S_{\mathrm{bond}} \) matrices. The resulting ABS is shown
in \cref{fig:edge-newtork} (b)--(d). In \cref{fig:edge-newtork}
(b) for \( P_f = 0.1, \gamma = \pi / 2 \), we are in the vicinity of the
pseudo-Landau level regime. We see that apart from two gapped flat ABS, 
two additional gapless (at \( \phi = \pi \)) dispersive ABS levels emerge because of the edges.
In this regime, the level crossings of the flat and dispersive edges suggest
that the bulk and edges of the network are decoupled. Note that the energy of
the flat ABS are different from that of the bulk (see Fig.~\ref{fig:abs} (d)).
This is a consequence of the finite size of the network in the transverse
direction. As the value of \( P_f \) is increased while keeping \( \gamma = \pi
/ 2 \) fixed (refer to Fig.~\ref{fig:edge-newtork} (b), (c)), new ABS emerge
with a narrow bandwidth (similar to the bulk ABS spectrum, refer to Fig.~\ref{fig:abs} (h), (l)). However, the gapless dispersive ABS still persists.  These ABS arise because of the edges and are responsible for the large \( 4\pi \) periodic Josephson current, as discussed below.

We compute the amplitude of incoming modes to demonstrate that the states correspond to gapless dispersive Andreev bound states localized at the network's edges. The incoming mode ($\bm a$) for an ABS $\epsilon$ at superconducting phase difference $\phi$ belongs to the null-space of the operator $(\mathbb I - S_{\text{node}}S_{\text{bond}}(\epsilon,\phi))$, i.e.,
\begin{align}
    [\mathbb I - S_{\text{node}}S_{\text{bond}}(\epsilon,\phi)]\bm a = 0
\end{align}
From the solution, the amplitude of the incoming mode ($A_{n,m}$) at $(n,m)$-th node is given by
\begin{align}
    A_{n,m} = \sum_{i=1}^3 \sum_{\eta=1}^{2} \sum_{\xi\in\{\K,\Kp\}} 
    \left[\bm a_{i\xi}^{(n,m;\eta)}\right]^\dagger
    \bm a_{i\xi}^{(n,m;\eta)}.
    \label{eq:ampl}
\end{align}
$A_{n,m}$ is normalized, i.e.,  $\sum_{n,m}A_{n,m}=1$.
The amplitudes of the nodes of the network are presented in \cref{fig:edge-state}. For $P_f=0.1$ and $\gamma=\pi/2$ (as illustrated in \cref{fig:edge-newtork} (b)), we depict the amplitude for the positive and negative gapless dispersive ABS in Fig.~\ref{fig:edge-state} (a) and (b), respectively, with $\phi=0$. These amplitudes are localized at the top or the bottom edge of the network and decay in the bulk. This confirms that these ABS are indeed edge states.

As there is no additional summation over Bloch momentum, the Josephson current is now computed as:
\begin{align}
	J(\phi) &= -k_B T \frac{2 e}{\hbar} \dv{}{\phi}\sum_{p = 0}^{\infty}
			\ln \det[1 - S_{\text{node}} S_{\text{bond}}(i\Omega_p,\phi)].
\end{align}
In Fig.~\ref{fig:edge-newtork} (e), we see that near the pseudo-Landau level
for \( P_f = 0.1, \gamma = \pi / 2 \) in the presence of the top and bottom
edges, the current is finite, where it was vanishingly small for the bulk Josephson junction (please refer to Fig.~\ref{fig:Josephson current} (a)). The Josephson current has a \( 4 \pi \) periodic
nature. As we increase the value of \( P_f \) while keeping \( \gamma = \pi / 2
\) in Fig.~\ref{fig:edge-newtork} (f), the critical current increases as a
result of increased forward scattering while retaining the \( 4\pi \) periodic nature 
of the Josephson current. For larger values of \( P_f \), the Josephson current
also attains a \( 2 \pi \) periodic component, thus making the Josephson current
skewed as a function of \( \phi \).
\section{Discussion and summary}
\label{sec:discussion}
In this study, we have explored the Josephson junction comprising of a minimally
twisted bilayer graphene sandwiched between two $s$-wave
superconductors. Employing a Chalker-Coddington network model of the minimally
twisted bilayer graphene, consistent with microscopic symmetries \cite{de_beule_aharonov-bohm_2020}, we investigate the system's transport phenomena across various network parameters. Depending on the network parameters of the minimally twisted bilayer graphene, the system exhibits distinct Andreev bound states and Josephson currents. Specifically, for zig-zag modes, we observe the emergence
of zero-energy Andreev bound states when the superconducting phase difference
is $\pi$, leading to a $4\pi$ periodic Josephson current with a sawtooth profile. Conversely, in the case of pseudo-Landau level modes, the Andreev bound states manifests as a perfectly flat spectrum, resulting in the vanishing of the Josephson current when periodic boundary condition is assumed in the transverse direction. However, when edges are present in the network, the pseudo-Landau level modes can mediate a \( 4\pi \) periodic  Josephson current through those edges even if the Josephson current in the bulk vanishes. This is similar to a skipping orbit of electrons transporting through the edge of a 2D electron gas in a magnetic field.
Additionally, depending on the network parameters, the Josephson current may
exhibit $2\pi$ periodicity, $4\pi$ periodicity, or a combined character of both. 

The $4\pi$-periodic Josephson current has several distinct experimental signatures. Firstly, under constant DC bias ($V$), the oscillating Josephson current shows a dipolar Josephson emission at half of the Josephson frequency $f_J/2 = e V / h$, typically in the GHz range, which can be measured using RF techniques \cite{Bocquillon2018}. Secondly, in the presence of an external microwave excitation at frequency $f$, Shapiro steps appear at discrete voltages given by $V_n = n h f / 2 e$, where $n$ is an integer step index. In the presence of a sizable $4\pi$-periodic Josephson current, only even
steps (with missing odd steps) are expected \cite{Park_2021,Wiedenmann_2016}. Additionally, bolometric detection of Josephson radiation may also reveal this $4\pi$ periodicity \cite{karimi2024bolometric}.

In our study, we have adopted a phenomenological approach, as the microscopic origins of $P_f$ and $\gamma$ remain unknown. In general, it may depend on the particulars of device fabrication, the value of the electric field, the twist angle, substrate potential, and other factors.
From a physical perspective, besides the interlayer bias, the presence of a periodic potential, applied to the scattering nodes, can induce repulsion for the propagating domain-wall electrons, thereby effectively reducing the amplitude of forward scattering $P_f$. Tuning $\gamma$ is more challenging to achieve. One could apply a staggered magnetic field to append an additional phase to the domain wall modes, thus altering \(\gamma\) while maintaining time reversal symmetry in the system. Achieving such control in practice would enable tuning of \( P_f \) and \( \gamma \), thereby allowing us to modify the character of the Josephson junction and produce \( 2\pi \) and \( 4\pi \) periodic Josephson currents on demand, in situ.

Our study can also be adapted to recently discovered moiré systems, such as helical
trilayer graphene \cite{devakul2023magic}. Like the minimally
twisted bilayer graphene, this system also
forms triangular domains due to lattice relaxation. In contrast to the minimally
twisted bilayer graphene,
helical trilayer graphene domains do not demonstrate AB/BA Bernal stacking;
rather, each domain possesses a single-moiré periodic structure that is
connected via $C_{2z}$ symmetry. If these domains can be gapped,
electronic transport is primarily governed by domain-wall modes.

\begin{acknowledgments}
R.K. and A.K. acknowledge Diptiman Sen, Mandar M. Deshmukh, Ganapathy Murthy, and Ankur Das for illuminating discussions. R.K. acknowledges funding under the PMRF scheme (Govt. of India). A.K. acknowledges support from the SERB (Government of India) via Sanction No. ECR/2018/001443 and CRG/2020/001803, DAE (Government of India) via Sanction No. 58/20/15/2019-BRNS, and MHRD (Government of India) via Sanction No. SPARC/2018-2019/P538/SL.
\end{acknowledgments}

\bibliography{refs2000.bib}

\appendix

\setcounter{figure}{0}
\renewcommand{\thefigure}{S\arabic{figure}}

\section{Symmetry constraints on Scattering matrix}
\label{sec:app:symm}
Following Ref. \cite{Roy2017}, we derive the action of symmetry operations on the scattering matrix. Below we only assume spin-less symmetries as minimally twisted bilayer graphene has spin-rotation symmetry intact.
\subsection{Time reversal operation}
In the spinless case, the time reversal operation is simply complex conjugation: \( \mathcal{T} = \mathcal{K} \), where \( \mathcal{K} \) denotes complex conjugation. Time reversal also interchanges the incoming and outgoing modes, with the action given by:
\begin{align}
    \label{eq:app:trs1}
    &\mathcal{T} \bm b_{\text{out}} = \bm a^*_{\text{in}},\quad
    \mathcal{T} \bm a_{\text{in}} = \bm b^*_{\text{out}}, 
\end{align}
where \( \bm b_{\text{out}} \) (\( \bm a_{\text{in}} \)) is the vector consisting of outgoing (incoming) modes.

Let \( S_{+} \) and \( S_{-} \) denote the scattering matrices for the time-forward and time-reversed processes, related by the \( \mathcal{T} \) operator as:
\begin{align}
    \label{eq:app:sminus}
    S_{-} = \mathcal{T} S_{+} \mathcal{T}^{-1}.
\end{align}
For \( S_+ \), we have:
\begin{align}
    \label{eq:app:scat1}
    \bm b_{\text{out}} = S_{+} \bm a_{\text{in}}.
\end{align}
Acting the time reversal operation on this equation, we obtain:
\begin{align}
    \label{eq:app:scat2}
    &\mathcal{T} \bm b_{\text{out}} = \mathcal{T} S_{+} \mathcal{T}^{-1} \mathcal{T}
    \bm a_{\text{in}}, \\
    \Rightarrow &\bm a^*_{\text{in}} = S_{-} \bm b^*_{\text{out}}, \\
    \label{eq:app:scat3}
    \Rightarrow &\bm b_{\text{out}} = S^{t}_{-} \bm a_{\text{in}}. 
\end{align}
Here, \( S^{t}_{-} \) denotes the transpose of \( S_{-} \). To arrive at \cref{eq:app:scat3}, we utilize \cref{eq:app:sminus,eq:app:trs1}. 

Thus, for time-reversal invariant systems, if we know the time-forward scattering matrix \( S_+ \), then \( S_{-} = S^{t}_{+} \).

Moreover, if the scattering matrix is energy and momentum-dependent, after time reversal, the energy argument remains the same while the momentum changes sign. Hence, \( S_{-}(E,k) = S^{t}_{+}(E,-k) \).

\subsection{Charge conjugation operation}
Charge conjugation interchanges particles and holes with a conjugation, i.e.,
\( \mathcal C = \tau_x \mathcal K \), here, \( \tau_x \) is the Pauli matrix acting on the particle-hole
index. The action of 
charge conjugation operation on incoming and outgoing modes is given by:
\begin{align}
	\label{eq:app:ph}
	&\mathcal C \bm b_{\text{out}} = \tau_x  \bm b^*_{\text{out}},\quad
	\mathcal C \bm a_{\text{in}} = \tau_x \bm a^*_{\text{in}}.
\end{align}
Let the scattering matrix of particle and hole sectors be denoted by \( S_p \) and
\( S_h \), respectively. They are related by the charge conjugation operator:
\begin{align}
	S_h = \mathcal C S_p \mathcal C^{-1}.
\end{align}
For \( S_p \) we have 
\begin{align}
	\label{eq:app:scat-ph1}
	\bm b_{\text{out}} = S_p \bm a_{\text{in}}.
\end{align}
Upon applying charge conjugation to both sides of 
\cref{eq:app:scat-ph1},
we obtain:
\begin{align}
	&\mathcal C \bm b_{\text{out}} = \mathcal C S_p \mathcal C^{-1} \mathcal C
	\bm a_{\text{in}}, \\
	\Rightarrow&\tau_x \bm b^*_{\text{out}} = S_h \tau_x \bm a^*_{\text{in}}, \\
	\label{eq:app:scat-ph2}
	\Rightarrow&\bm b_{\text{out}} =\tau_x S^*_h \tau_x \bm a_{\text{in}}.
\end{align}
Together \cref{eq:app:scat-ph1,eq:app:scat-ph2} imply \( S_h =
\tau_x S^*_p \tau_x\). For the case studied in the main text, the scattering
matrix is for a normal metallic region diagonal in the particle-hole index.
So, for that case, charge conjugation symmetry implies \( S_h = S_p^* \). 

In the case of an energy and momentum-dependent scattering matrix, charge conjugation flips the sign of both energy and momentum, yielding \( S_h(E,k) = S_p^*(-E,-k) \).

\section{Derivation of \cref{eq:Sk}}
\label{Sk derivation}

\begin{figure}[ht]
	\centering
	\includegraphics{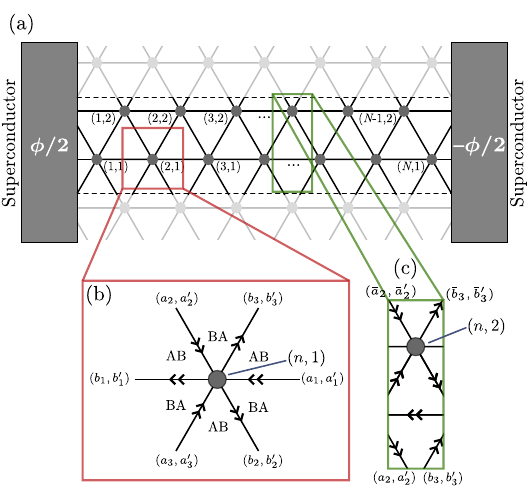}
	\caption{
		(a) The geometry considered here for the Josephson junction, composed of two $s$-wave superconducting leads (gray rectangles) connected by MTBG. Nodes are indexed by two integers $(m,n)$, which are connected by network links (black lines). We assume a periodic boundary condition in the transverse direction. The dashed lines encapsulate the unit-cell of the network.
		(b) The indexing of modes for a node with index \( (n,1) \). 
		(c) The indexing of modes for a node with index \( (n,2) \) used to derive \cref{eq1,eq2,eq3,eq4}. 
	(b) and (c) are for K-valley. For K$'$-valley the directions of the modes are interchanged, i.e., \( a \leftrightarrow b \).}
	\label{fig:node-modes}
\end{figure}
In this section we show the derivation of \cref{eq:Sk}.
For this purpose we only consider the node indices and drop all other indices, as the following
identities hold for all the flavor indices. The scattering matrix acts on incoming modes
and returns the outgoing modes at each index. 

For the nodes with index \( (n,1) \), we have:
\begin{align}
	\label{eq:noden1}
	\left(b_{1}, b_{2}, b_{3}, b_{1}', b_{2}', b_{3}'\right)^T
	=  S
	\left(a_{1}, a_{2}, a_{3}, a_{1}', a_{2}', a_{3}'\right)^T
\end{align}
\( S \) is defined in \cref{eq:smat1} of the main text.

Special care has to be taken to write done similar relations for the nodes \(
(n,2) \) as the network model enjoys periodicity in the transverse direction and modes 
are not independent at the top and bottom of the unit cell (see to \cref{fig:node-modes} (c)).

\begin{align}
	\left(b_{1}, b_{2}, \bar b_{3}, b_{1}', b_{2}', \bar b_{3}'\right)^T
	=  S
	\left(a_{1}, \bar a_{2}, a_{3}, a_{1}', \bar a_{2}', a_{3}'\right)^T
\end{align}
Here, \( \bar a_2, \bar a_2' \) and \( \bar b_3, \bar b_3' \) denote
inter-unit-cell incoming modes, and outgoing modes at the top edge of the unit
cell, respectively. The subscript indicate the direction of the modes as shown
in \cref{fig:node-modes}. Now, we invoke the Bloch's theorem (see \cref{fig:node-modes} (c))
\begin{align}
	\label{eq1}
	\bar b_3 &= e^{i \sqrt 3 k l} b_3, \\
	\label{eq2}
	\bar b_3' &= e^{i \sqrt 3 k l} b_3', \\
	\label{eq3}
	\bar a_2 &= e^{i \sqrt 3 k l} a_2, \\
	\label{eq4}
	\bar a_2' &= e^{i \sqrt 3 k l} a_2'.
\end{align}
Here \( k \in [0,2\pi / \sqrt 3 l)\).
This results in:
\begin{widetext}
\begin{align}
	&\left(b^{}_{1}, b^{}_{2}, e^{i \sqrt3 k l}b^{}_{3}, b'_{1}, b'_{2}, e^{i \sqrt3 k l} b'_{3}\right)^T
	= S
	\left(a^{}_{1}, e^{i \sqrt 3 k l} a^{}_{2}, a^{}_{3}, a'_{1}, e^{i \sqrt 3 k l} a'_{2}, a'_{3}\right)^T,\\
	&\Rightarrow U_1(k)^{-1} \left(b_{1}, b_{2}, b_{3}, b_{1}', b_{2}', b_{3}'\right)^T = S  U_2(k) \left(a_{1}, a_{2}, a_{3}, a_{1}', a_{2}', a_{3}'\right)^T, \\
	&\Rightarrow \left(b_{1}, b_{2}, b_{3}, b_{1}', b_{2}', b_{3}'\right)^T = U_1(k) S  U_2(k) \left(a_{1},  a_{2}, a_{3}, a_{1}',  a_{2}', a_{3}'\right)^T,
\end{align}
\end{widetext}
where \( U_1(k) = \mathbb I_2 \otimes \text{diag}(1,1,e^{-i \sqrt 3 k l}) \) and 
\( U_2(k) = \mathbb I_2 \otimes \text{diag}(1,e^{i \sqrt 3 k l},1) \). This proves \cref{eq:Sk} of the main text.

\section{Details of the \texorpdfstring{\(S_{\text{bond}}\)}{bond} matrix}
\label{sec:app:sbond}
This section details the construction of the $S_{\text{node}}$ matrix. Following each scattering event, the outgoing modes propagate freely for a
duration of \( \tau = l / v_F \), ($l$ is the moiré length scale and $v_F$ is the velocity of DW modes) before the subsequent scattering event. During
this time, the DW modes acquire a dynamic phase \( \xi \epsilon l/ \hbar v_F \), where \(
\epsilon \) is the energy of the propagating modes. The valley index ($\xi$) in the
dynamical phase accounts for the DW modes propagating in opposite
directions in different valleys.
After time \( \tau \), we write down all the connections between the outgoing and incoming
modes between the neighboring nodes. For K-valley, we have:
 \begin{align}
	 \label{eq:link-first}
	 \bm a_{3\K}^{(n,2;\eta)} &= e^{i \epsilon l / v_F}    \bm b_{3\K}^{(n,1;\eta)}, \\
	 \bm a_{2\K}^{(n+1,1;\eta)} &= e^{i \epsilon l / v_F}  \bm b_{2\K}^{(n,2;\eta)}, \\
	 \bm a_{1\K}^{(n,1;\eta)} &= e^{i \epsilon l / v_F }   \bm b_{1\K}^{(n+1,1;\eta)}, \\
	 \bm a_{1\K}^{(n,2;\eta)} &= e^{i \epsilon l / v_F}    \bm b_{1\K}^{(n+1,2;\eta)}, \\
	 \bm a_{3\K}^{(n+1,1;\eta)} &= e^{i \epsilon l /v_F }  \bm b_{3\K}^{(n,2;\eta)}, \\
	 \bm a_{2\K}^{(n,2;\eta)} &= e^{i \epsilon l / v_F}    \bm b_{2\K}^{(n,1;\eta)}.
 \end{align}
 Here, \( i \in \{1,2,3\} \) in the subscript of the modes \(\{i \xi\}\) refers
 to the three incoming and outgoing channels, \( \eta \in \{1,2\} \) runs
 over two DW modes per valley, per spin. Here, \( \bm a_{i\xi}^{(n,m;\eta)} \) and  \( \bm b_{i\xi}^{(n,m;\eta)} \) represent the amplitudes of the incoming and outgoing modes by four-dimensional vectors written in the basis of  \( \{e,h\} \otimes \{\uparrow,\downarrow\}\). 
 For K\('\)-valley, the modes traverse in the opposite direction,
 resulting in a change in the sign of velocity (\(v \to -v\)). In other words,
 the incoming and outgoing modes interchange (\(a \leftrightarrow b\)) on the K$'$-valley. The connections between outgoing and incoming modes between the neighboring nodes in the K\('\)-valley can be expressed as follows:
\begin{align}
	\bm a_{3\Kp}^{(n,1;  \eta)}&= e^{i \epsilon l / v_F}   \bm b_{3\Kp}^{(n,2;  \eta)}, \\
	\bm a_{2\Kp}^{(n,2;  \eta)} &= e^{i \epsilon l / v_F}  \bm b_{2\Kp}^{(n+1,1;\eta)},  \\
	\bm a_{1\Kp}^{(n+1,1;\eta)} &= e^{i \epsilon l / v_F } \bm b_{1\Kp}^{(n,1; \eta)}, \\
	\bm a_{1\Kp}^{(n+1,2;\eta)}&= e^{i \epsilon l / v_F}   \bm b_{1\Kp}^{(n,2; \eta)},\\
	\bm a_{3\Kp}^{(n,2;  \eta)}&= e^{i \epsilon l /v_F }   \bm b_{3\Kp}^{(n+1,1;\eta)},\\
	\bm a_{2\Kp}^{(n,1;  \eta)}&= e^{i \epsilon l / v_F}   \bm b_{2\Kp}^{(n,2;  \eta)}.
	 \label{eq:link-last}
\end{align}

At the superconducting electrodes, an electron of \( K \) valley and \( \uparrow \)-spin is Andreev reflected as a hole of \( K' \) valley with \( \downarrow \)-spin and vice versa. 
As the links are made from one-dimensional DW modes, only retro Andreev reflection takes place,
and specular Andreev reflection is suppressed \cite{beenakker_specular_2006}.
The Andreev reflection is incorporated in the left and right lead via the following matrix:
\begin{align}
	M_{A}(\phi) = i \alpha
	\begin{pmatrix}
		0 & 0 & 0 & -e^{i \phi} \\
		0 & 0 & e^{i \phi} & 0 \\
		0 & e^{-i \phi} & 0  & 0 \\
		-e^{-i \phi} & 0 & 0 & 0
	\end{pmatrix}.
\end{align}
\( \alpha \) is defined as,
\begin{align}
	\alpha = 
	\begin{cases}
		i \exp(- i \cos^{-1}\left({\epsilon}/{\Delta}\right))\quad \text{for } \epsilon \le  \Delta \\
		i \exp(- \cosh^{-1}\left({\epsilon}/{\Delta}\right)) \quad \text{for } \epsilon > \Delta,
	\end{cases}
\end{align}
where $\Delta$ and $\phi$ are the superconducting gap and phase, respectively. The Andreev reflection connects outgoing and incoming modes on the left
superconducting junction as follows:
\begin{align}
	\label{eq:link-first-andreev}
	\bm a_{1\K}^{(1,2; \eta)} &= M_A(\phi/2) \bm b_{1 \Kp}^{(1,2;\eta)},\\
	\bm a_{2\Kp}^{(1,1;\eta)} &= M_A(\phi/2) \bm b_{2 \K}^{(1,1; \eta)},\\
	\bm a_{1\K}^{(1,1; \eta)} &= M_A(\phi/2) \bm b_{1 \Kp}^{(1,1;\eta)},\\
	\bm a_{3\Kp}^{(1,1;\eta)} &= M_A(\phi/2) \bm b_{3 \K}^{(1,1; \eta)}.
\end{align}
Similarly, the connections at the right superconducting lead are given by:
\begin{align}
	\bm a_{1\Kp}^{(N-1,2;\eta)} &= M_A(-\phi/2) \bm b_{1 \K}^{(N-1,2;\eta)},\\
	\bm a_{2\K}^{(N,1;   \eta)} &= M_A(-\phi/2) \bm b_{2 \Kp}^{(N,1; \eta)},\\
	\bm a_{1\Kp}^{(N,1;  \eta)} &= M_A(-\phi/2) \bm b_{1 \K}^{(N,1;  \eta)},\\
	\bm a_{3\K}^{(N,1;   \eta)} &= M_A(-\phi/2) \bm b_{3 \Kp}^{(N,1; \eta)}.
	\label{eq:link-last-andreev}
\end{align}
The equations \eqref{eq:link-first}--\eqref{eq:link-last} for the MTBG links
and \eqref{eq:link-first-andreev}--\eqref{eq:link-last-andreev} for the
superconducting leads and normal MTBG junction define the bond matrix \(
S_{\text{bond}} \) that acts on outgoing modes and returns incoming modes,
i.e., $$\bm a = S_{\text{bond}} \bm b.$$

\section{Extended data for Andreev bound states}
\label{app:Extended data for Andreev bound states}
\begin{figure*}[htpb]
	\centering
	\includegraphics[width=500pt]{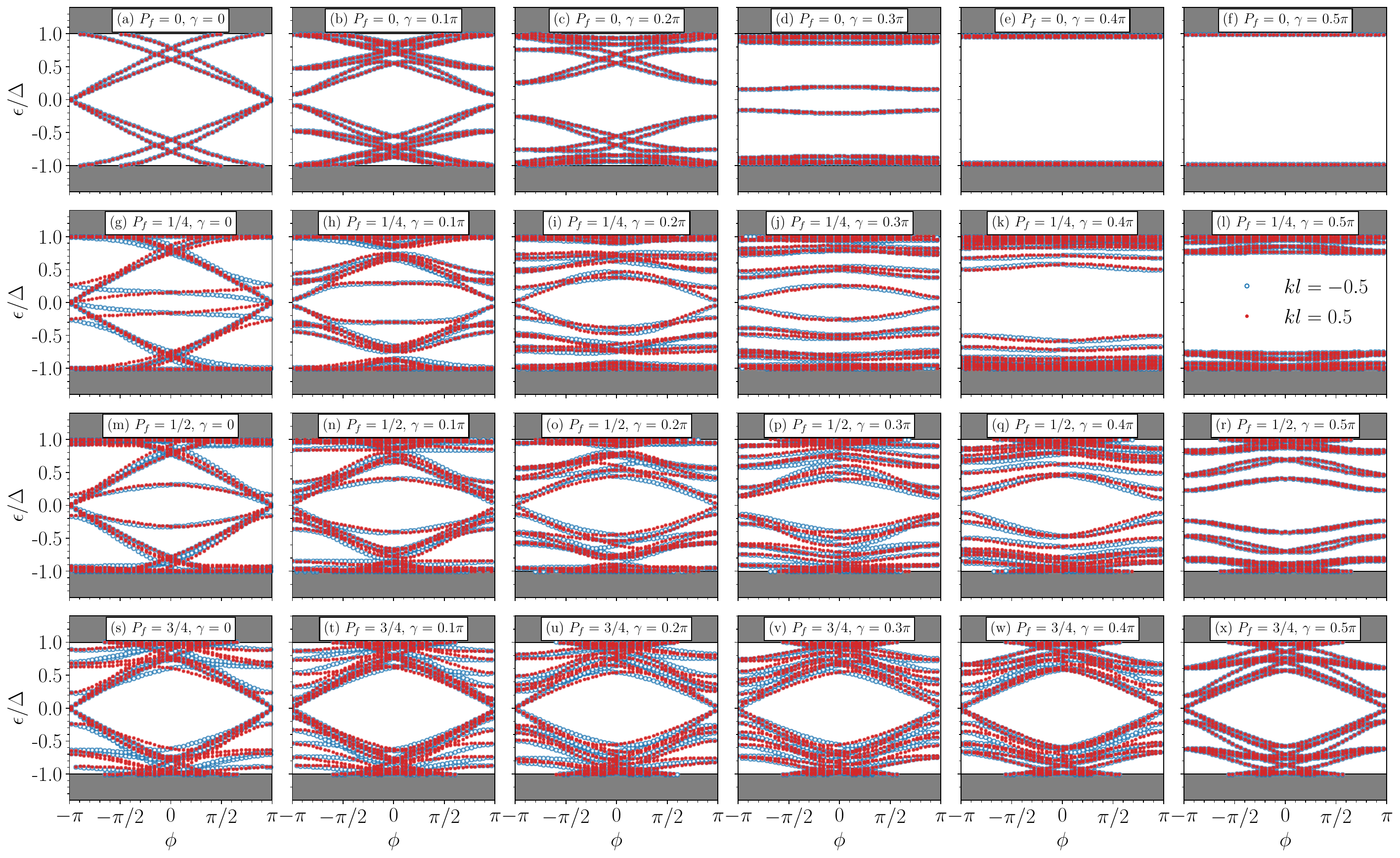}
	\caption{Extended \cref{fig:abs} of main text.
 Andreev bound states (ABS) with different choice of \( P_f \) and
		\( \gamma\) and for two representative values of transverse momentum \(
		k l = 0.5 \) (red) and \(k l = -0.5\) (blue). The gray regions denote the continuum of states \( \abs{\epsilon} >
		\Delta \). We use $\theta=0.1^\circ$, $N=3$, $\Delta=1$~meV for these numerical plots.
 }
	\label{fig:abs-ex}
\end{figure*}

In this section, we present \cref{fig:abs-ex}, extended data of the Andreev bound state shown in the \cref{fig:abs} for other values of network parameters.

\section{Asymmetric choice of parameters}
\label{sec:asym}
\begin{figure*}[htpb]
	\centering
	\includegraphics[width=450pt]{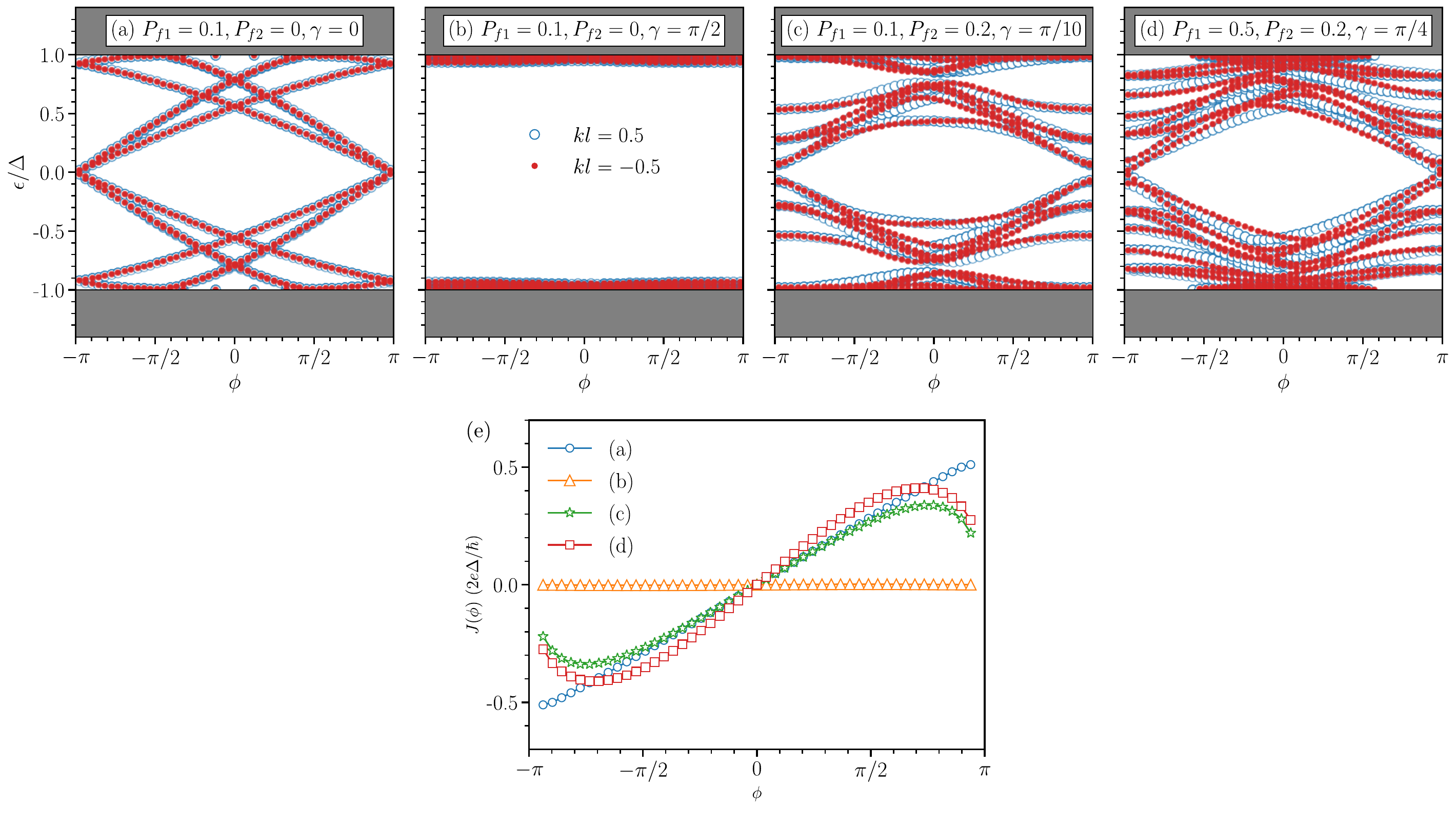}
	\caption{
(a)--(d) show the ABS with \( P_{f1} \neq P_{f2} \) and \( P_{d1} = P_{d2} \) for representative network parameter values. The blue circles represent Bloch momentum \( kl = 0.5 \), and the red disks represent \( kl = -0.5 \). Figure (e) shows the corresponding Josephson current. For these numerical plots, we use \( \theta = 0.1^\circ \), \( N = 3 \), \( \Delta = 1 \)~meV, and \( T = 0.01 \Delta \). The numerical plots are qualitatively similar to \cref{fig:abs} and \cref{fig:Josephson current} (a), (b) for symmetric choice of parameters \( P_{f1} = P_{f2} \) .
 }
	\label{fig:abs-pf12-pdf}
\end{figure*}

Throughout the main text we have assumed the symmetric choice of network parameters
for simplicity, i.e., \( P_{f1} = P_{f2} \) and \( P_{d1} = P_{d2} \). Here,
we present some results when \( P_{f1} \neq P_{f2} \) and retain \( P_{d1} = P_{d1} \).
\cref{fig:abs-pf12-pdf} summarizes the numerical results of ABS and Josephson current
for such choice of parameters.

Near the near the zig-zag mode parameters with this asymmetry, i.e., \( P_{f1}
= 0.1, P_{f2} = 0, \gamma = 0 \) as shown in \cref{fig:abs-pf12-pdf} (a), the
ABS is qualitatively similar to those of the symmetric parameters (see
\cref{fig:abs} (a)). Near the PLL modes, i.e., \( P_{f1} = 0.1, P_{f2} = 0,
\gamma = \pi / 2 \), as shown in \cref{fig:abs-pf12-pdf} (b), the ABS becomes
nearly flat (as seen for symmetric parameters in \cref{fig:abs} (d)). We show
for two more values of network parameters in \cref{fig:abs-pf12-pdf} (c), (d).
The corresponding Josephson current is shown in \cref{fig:abs-pf12-pdf} (e).

\end{document}